\documentclass[journal]{IEEEtran}
\usepackage{xcolor,soul,framed} 
\colorlet{shadecolor}{yellow}
\usepackage[pdftex]{graphicx}
\graphicspath{{../pdf/}{../jpeg/}}
\DeclareGraphicsExtensions{.pdf,.jpeg,.png}
\usepackage[cmex10]{amsmath}
\usepackage{array}
\usepackage{mdwmath}
\usepackage{mdwtab}
\usepackage{eqparbox}
\usepackage{url}
\usepackage{subfigure}
\usepackage{graphicx}

\begin{document}
\bstctlcite{IEEEexample:BSTcontrol}
    \title{An Analytical and Rigorous Method for Analysis of an Array of Magnetically-Biased Graphene Ribbons}
  \author{Mahdi~Rahmanzadeh,~Behzad~Rejaei,~Mohammad Memarian,~\IEEEmembership{Senior Member,~IEEE}, and Amin~Khavasi

  \thanks{The authors are with the Department of Electrical Engineering, Sharif University of Technology, Tehran 11155-4363, Iran (email:rahmanzadeh.mahdi@ee.sharif.edu; rejaei@sharif.edu; mmemarian@sharif.edu (Corresponding author is M. Memarian); khavasi@sharif.edu}} 

\markboth{}{Rahmanzadeh\MakeLowercase{\textit{et al.}}:An Analytical and Rigorous Method for Analysis of an Array of Magnetically-Biased Graphene Ribbons}

\maketitle

\begin{abstract}
A sheet of graphene under magnetic bias attains  anisotropic surface conductivity, opening the door for realizing compact devices such as Faraday rotators, isolators and circulators. In this paper, an accurate and analytical method is proposed for a periodic array of graphene ribbons under magnetic bias. The method is based on integral equations governing the induced surface currents on the  coplanar array of graphene ribbons. For subwavelength size ribbons subjected to normally incident plane waves, the current distribution is derived leading to analytical expressions for the reflection/transmission coefficients. The results obtained are in excellent agreement with full-wave simulations and predict resonant spectral effects that cannot be accounted for by existing semi-analytical methods. Finally, we extract an analytical, closed form solution for the Faraday rotation of magnetically-biased graphene ribbons. In contrast to previous studies, this paper presents a fast, precise and reliable technique for analyzing magnetically-biased array of graphene ribbons, which are one of the most popular graphene-based structures.
\end{abstract}

\IEEEpeerreviewmaketitle
\IEEEpeerreviewmaketitle

\section{Introduction}
\IEEEPARstart{G}{raphene} is a two-dimensional (2D) nanomaterial that is composed of a single layer of carbon atoms organized into a hexagonal lattice \cite{geim2010rise}. In the past decade, graphene-based devices have become a prime focus of research due to their unique electrical, optical, thermal, and mechanical properties  \cite{novoselov2004electric},\cite{balandin2008superior},\cite{bao2012graphene}. The electrical conductivity of graphene can change by an external DC voltage or chemical doping. This tunability provides additional flexibility in the design of diverse functionalities. Therefore, numerous graphene-based applications have been proposed from microwave to terahertz regions\cite{rahmanzadeh2018multilayer,momeni2018information,chen2013nanostructured,rahmanzadeh2017analytical,abdollahramezani2015analog,khavasi2015design,filter2013tunable,rouhi2019multi} Under an external magnetic bias, the surface conductivity of graphene becomes a tensor, and it will also have non-reciprocal properties \cite{hanson2008dyadic}. These extraordinary properties make graphene a promising material for many novel applications, such as isolators \cite{tamagnone2016near}, nonreciprocal couplers \cite{chamanara2013non}, optical modulators \cite{tamagnone2014fundamental}, and Faraday rotators \cite{crassee2011giant}, in the terahertz regime.
\par Due to strong interaction of graphene with electromagnetic (EM) fields, patterned magnetically-biased graphene has been assessed up to now for several applications \cite{poumirol2017electrically,fallahi2012manipulation,tymchenko2013faraday}. For example, in \cite{tymchenko2013faraday} it has been shown that the Faraday rotation of graphene micro ribbons can be controlled dynamically and that a giant Faraday rotation can be obtained, the frequency of which corresponds to the width of ribbons.
\par Consequently, the analytical and numerical study of continuous and patterned anisotropic graphene is currently of high interest. A continuous graphene sheet can be easily analyzed by applying appropriate boundary conditions. Saunas and Caloz studied the transmission and reflection properties of magnetically-biased graphene sheet using the general anisotropic conductivity tensor \cite{sounas2011electromagnetic}.
To the best of the authors' knowledge, no rigorous and analytical solution has yet been presented on patterned anisotropic graphene under magnetic bias. Although certain numerical methods, including the Fourier modal-based \cite{tymchenko2013faraday}, the finite difference time-domain method \cite{wang2013matrix}, and discontinuous Galerkin time domain (DGTD) \cite{li2015modeling} have been used to numerically solve this problem, they suffer from a slow convergence rate and more importantly, are not analytic results. Furthermore, some efforts have been made to decrease the computation time \cite{khoozani2016analysis,feizi2018modeling}; however, they still require higher acceleration. In addition, a simple method for the analysis of magnetic-biased graphene ribbons based on the effective medium approach was proposed in  \cite{gomez2016magnetically}; however, as it will be shown, this method does not have accuracy and is only valid for the extreme subwavelength regime. Hence a rigorous and analytical analysis of magnetically-biased Graphene-based meta-surfaces is of paramount importance.
\par In this paper, a rigorous analysis is performed through solving the integral equations governing the induced surface currents on the graphene ribbons to accurately predict the electromagnetic performance of an array of magnetically-biased graphene ribbons. The paper is organized as follows: In Sec. II we shall first study the scattering of a TE/TM electromagnetic wave by a single magnetically-biased graphene ribbon. Then  an analytical expression is extracted for the surface current density induced within a periodic array of magnetically-biased ribbons in the subwavelength regime. In Sec. III, the reflection/transmission coefficient of the structure is calculated and the proposed method is validated against full-wave simulations. The limitations of our proposed method is discussed. Moreover, an analytical expression is presented for Faraday rotation of magnetically-biased graphene ribbons, as one of the important phenomena in optics and photonics.

\section{SURFACE CURRENTS ON   A PERIODIC ARRAY OF ANISOTROPIC GRAPHENE RIBBONS}
We shall consider a periodic array of 1D graphene ribbons under magnetic bias as shown in Fig.\ref{GrapeheneRibbon}. Each ribbon has a width of $w$. The periodicity of the array is $D$ in the x-direction. The graphene ribbons are placed on the $x-y$ plane (infinite along $y$) and a static magnetic field $B_0$ is applied perpendicular to the ribbons. A normal EM plane wave is incident on the arrangement. Due to the incident wave, a surface current density with components $J_x$ and $J_y$ will be induced on the ribbons. Since there is no variation in  $y$-direction, these induced currents are functions of $x$ only  (i.e. $\partial_{y} {J_{x,y}}= 0$).

\subsection {Electromagnetic parameters of magnetically-biased graphene}

As noted earlier, graphene is a 2-D nanomaterial consisting of a single layer of carbon atoms organized within a hexagonal lattice. Assume that the graphene sheet is on the  $x-y$ plane and that a magnetic field is also established perpendicular to the graphene sheet ($\overrightarrow{B}={B_0 \hat{z}}$). The surface conductivity of this graphene sheet is represented by a tensor (${\overline{\overline \sigma } _s}$) that is obtained from microscopic, semi-classical, and quantum mechanical considerations as \cite{hanson2008dyadic}
\begin{equation}
\label{eq1}
{\overline{\overline \sigma } _s}(\omega ,{E_f},\tau ,T,{B_0})= \left[\begin{array}{*{20}{c}}{{\sigma _{xx}}}&{{\sigma _{xy}}}\\{{\sigma _{yx}}}&{{\sigma _{yy}}}\end{array}\right] 
\end{equation}
where $E_f$ is the  Fermi level energy and $\omega$ is the angular frequency, $\tau$ is the relaxation time, and $T$ is the Kelvin temperature ($T= 300 K$). For highly doped graphene, i.e. $E_f>>\hbar\omega$ and $E_f>>k_B T$ ($k_B$ = Boltzmann constant, $\hbar$ = reduced Planck’s constant), (\ref{eq1})  can be simplified by a Drude-like model as  \cite{ferreira2011faraday}
\begin{subequations}
\label{eq2}
  \begin{equation}
    \label{eq2-a}
   {\sigma _{xx}} = {\sigma _{yy}} = \frac{{{e^2}{E_f}\tau}}{{\pi {\hbar ^2}}}\frac{{1 + j\omega \tau }}{{{{({\omega _c}\tau )}^2} + {{(1 + j\omega \tau )}^2}}}
  \end{equation}
  
  \begin{equation}
  \label{eq2-b}
   {\sigma _{xy}} =  - {\sigma _{yx}} = \frac{{{e^2}{E_f}}\tau}{{\pi {\hbar ^2}}}\frac{{{\omega _c}\tau }}{{{{({\omega _c}\tau )}^2} + {{(1 + j\omega \tau )}^2}}}
  \end{equation}
\end{subequations}
where $\omega_c={e}{B_0}{f^2_F}/{E_f}$ is the cyclotron frequency with $v_f=10^6 m/s$ denoting the Fermi velocity. The relaxation time can be described by $\tau  = \mu \hbar \sqrt {{n_s}\pi } /e{v_F}$  where $\mu$ is the carrier mobility and ${n_s} = E_F^2/\pi {\hbar ^2}v_F^2$  is the carrier density \cite{sounas2011electromagnetic}. Carrier mobility is assumed to be within the range of  $0.1$ to $6$ $m^2/V$, which is practical and achievable for graphene with different substrates at room temperature \cite{ju2011graphene,dean2010boron}. It should also be noted that the time harmonic of the form $\exp(j{\omega}t)$ is assumed throughout this paper.

\subsection{Scattering from a single ribbon}

\begin{figure}
\centering
\includegraphics[width=\linewidth]{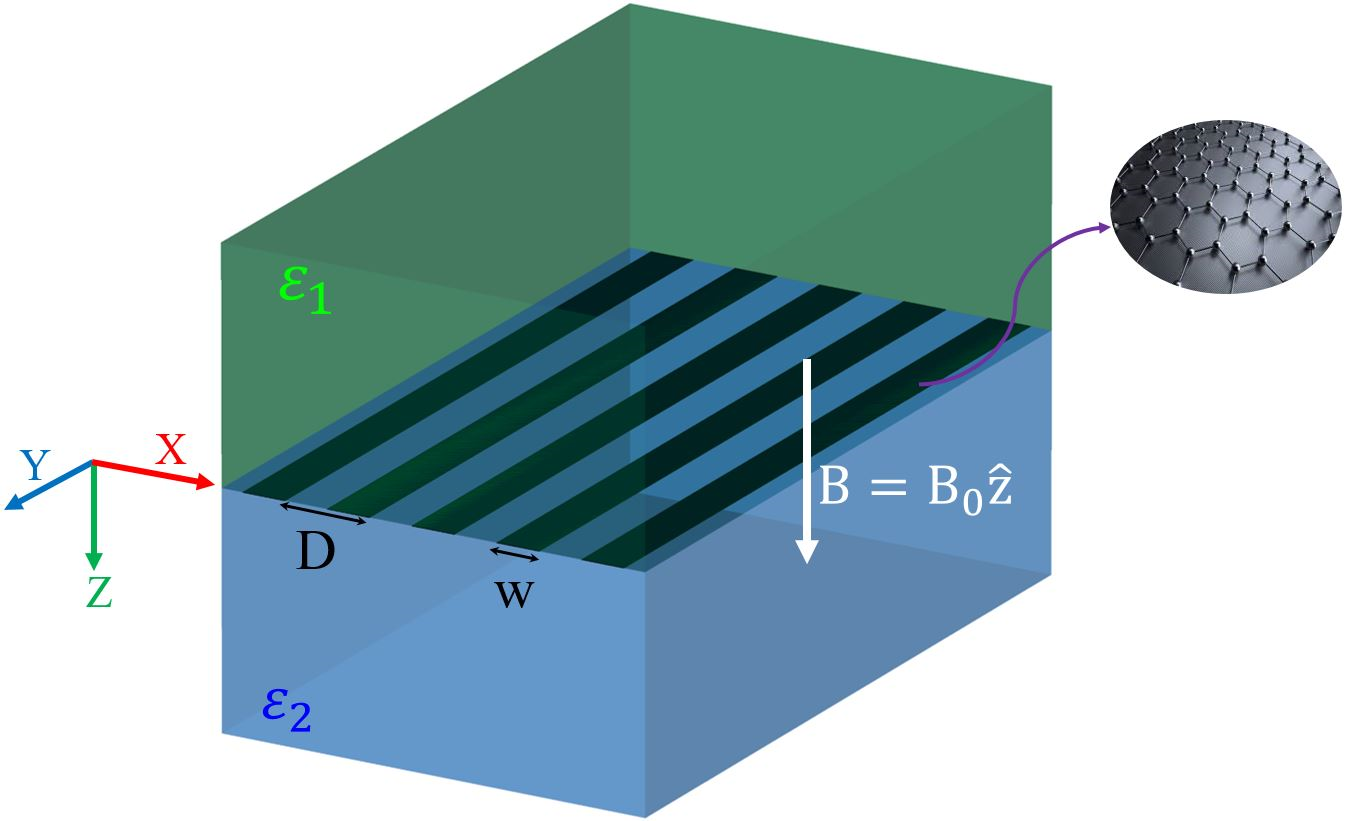}
\caption{Schematic representation of the studied system: a plane wave ($TE/TM$) is  incident on the array of graphene ribbons, in the presence of a static perpendicular magnetic field $B $.}\label{GrapeheneRibbon}
\end{figure}

Let us first consider a single anisotropic graphene ribbon in free space. The integral equation governing the induced surface currents on the single ribbon can be expressed as \cite{tai1994dyadic} 
\begin{subequations}
\label{eq3}
  \begin{equation}
  \label{eq3-a}
 {\overline{\overline {{\sigma _s}}}^{ - 1}}\left[\begin{array}{*{20}{c}}{{J_x}(x)}\\{{J_y}(x)}\end{array}\right] = \left[\begin{array}{*{20}{c}}{E_x^{ext}(x) + E_x^T(x)}\\{E_y^{ext}(x) + E_y^T(x)}\end{array}\right]
  \end{equation}
where $E_{x,y}^{ext}(x)$  denote $x$- and $y$-components of the incident (external) electric field, and
 \begin{align}
  \label{eq3-b}
\begin{split}
 &E_x^T = \frac{1}{j\omega \varepsilon _0} \frac{d}{dx} \int_{- w/2}^{w/2} G_0(x - x') \frac{dJ_x(x')}{dx'}dx' - \\
 & j\omega \mu _0\int\limits_{ - w/2}^{w/2} G_0(x - x') {J_x}(x')dx'
 \end{split}\\
\label{eq3-c}
&E_y^T =  - j\omega {\mu _0}\int\limits_{ - w/2}^{w/2} {{G_0}(x - x')} \,{J_y}(x')\,dx'
  \end{align}
\end{subequations}
represent the electric field generated by the surface current $J_x$ and $J_y$. In these equations
\begin{equation}
G_{0}(x-x')=\frac{1}{4j}H_{0}^{(2)}(k_{0}|x-x'|)\label{hank}
\end{equation}
designates the Green's function of the 2D Helmholtz equation. Here $H_{0}^{(2)}$ is the Hankel function of the second kind and $k_{0}=\omega (\varepsilon_{0}\mu_{0})^{1/2}$ is the vacuum wave number.

Assuming the strip to be narrow ($k_0w\ll 1$), we next adopt the quasi-static approximation \cite{khavasi2014analytical}. As a result, the second term on the right hand side of  \eqref{eq3-a} becomes negligible compared to the first term. Moreover, the Green's function \eqref{hank} can be approximated by
\begin{equation}
G_{0}(x-x')\simeq -\frac{1}{2\pi}\ln\left( k_{0}|x-x'|\right)\label{g0}
\end{equation}
From \eqref{eq3-a} and \eqref{eq3-c} one obtains
\begin{equation}
\label{eq4-b}
\begin{split}
J_y(x)+&j\omega\mu _0\sigma_{0}\int_{-w/2}^{w/2} G_{0}(x-x') \,J_y(x')\, dx' \\& = \sigma_{0}\left[ E_y^{ext}(x)-\sigma_{a}^{-1}  J_x(x)\right]
\end{split}
\end{equation}
where
\begin{equation}
\sigma_{0}^{-1}=\frac{\sigma_{xx}}{\sigma_{xx}^{2}+\sigma_{xy}^{2}},
\sigma_{a}^{-1}=\frac{\sigma_{xy}}{\sigma_{xx}^{2}+\sigma_{xy}^{2}}
\label{eq7}
\end{equation}
It must be noted that $\sigma_0$ is the graphene conductivity in the absence of the external magnetic field. 

Assume, for the moment, that $J(x)$ is known. \eqref{eq4-b} can then be viewed as can integral equation for the function $J_{y}(x)$. This integral equation has a complicated, but exact solution.  However, in the limit where
\begin{equation}\label{asy}
\left| \frac{j\omega\mu _0 w \sigma_{0}}{4}\right| \ll 1
\end{equation}
the second term on the left hand side of \eqref{eq4-b} becomes negligible and the solution is trivially given by  
\begin{equation}
\label{eq4-b-sol}
\begin{split}
J_y(x) = \sigma_{0}\left[ E_y^{ext}(x)-\sigma_{a}^{-1}  J_x(x)\right]
\end{split}
\end{equation}
To see why \eqref{asy} is satisfied, note that
\begin{equation}
\frac{j\omega\mu _0 w \sigma_{0}}{4}=(k_{0}w)\frac{k_{0}}{k_{P}}
\end{equation}

where $k_{P}=-2j\omega\varepsilon_{0}/\sigma_{0}$ is the wave number of graphene plasmons in the absence of an external magnetic field. Since $k_0 w\ll 1$ and $k_{0}\ll k_{P}$ at frequencies of interest, \eqref{asy} is satisfied. 

Let us now return to \eqref{eq3-a} and combine it with \eqref{eq3-b} to arrive at an equation for $J_{x}(x)$. However, one may simplify the resulting equation by neglecting the second term on the right hand side of \eqref{eq3-b}. This term is small compared to the first term for a narrow ribbon. The integral equation governing $J_{x}$ thus becomes
\begin{equation}\label{integ1}
\begin{split}
\frac{J_{x}(x)}{\sigma_{xx}}- &\frac{1}{j\omega \varepsilon _0} \frac{d}{dx} \int_{- w/2}^{w/2} G_0(x - x') \frac{d J_x(x')}{dx'}dx'\\=&E_x^{ext}(x)+\frac{\sigma_{xy}}{\sigma_{xx}}E_y^{ext}(x)
\end{split}
\end{equation}
The solution to this equation has been described in detail in  \cite{khavasi2014analytical}. The only difference with the work in  \cite{khavasi2014analytical} is the appearance of the second term on the right hand side of \eqref{integ1}.

\subsection{Scattering from an infinite array of ribbons}
Now consider a periodic array of coplanar graphene ribbons with period $D$ as shown in  Fig.\ref{GrapeheneRibbon}. The ribbon array is subject to a plane electromagnetic wave of arbitrary polarization that is normally incident on the array plane. Equation \eqref{eq3-a} that locally relates the ribbon surface currents to the total electric field is still applicable. However, on each ribbon, $E_{x,y}^{T}$ must now account for the total electric field that is produced by all ribbons. Fortunately, in case of normal incidence on an infinite number of array elements, the distributions of surface currents on all ribbons are identical. As a result, on each individual ribbon, \eqref{eq3-c} and \eqref{eq3-c} remain provided that the Green's function \eqref{hank} is replaced by
\begin{equation}
\label{phank}
G\left( x-x' \right) = \frac{1}{{4j}}\sum\limits_{l = -\infty}^{\infty}  {H_0^{(2)}\left( {{k_0}\left| {x - x' - lD} \right|} \right)} 
\end{equation}

As in the case of a single ribbon, we first treat the equation for $J_{y}$. However, the interaction between ribbons, manifested in the periodic Green's function \eqref{phank}, leads to an important modification of the approximation \eqref{g0}. For sub-wavelength arrays where $k_{0}D\ll 1$, a lengthy calculation presented in the appendix shows that   
\begin{equation}
\begin{split}
G(x)= \frac{1}{2jk_{0}D}+  \frac{1}{2\pi }
\int\limits_{0 }^{ \infty }\left[
\frac{\cosh\left( \alpha u \right) }{\sinh u } -\frac{1}{u}\right]\frac{du}{u} +O\left( k_{0}D\right)\label{gp}
\end{split}
\end{equation}
 The interaction between sub-wavelength ribbons leads to a Green's function whose leading term is of the order $\left( k_{0}D\right)^{-1}$. Unlike the case of a single ribbon, the second term in the integral equation \eqref{eq4-b} cannot be neglected anymore. By replacing  $G_{0}$ with \eqref{gp} in \eqref{eq4-b}, and keeping the leading term alone, one ends up with the equation
\begin{equation}
\label{eq4-ba}
J_y(x)=\sigma_{0}\left[ E_y^{ext}(x)-\sigma_{a}^{-1}  J_x(x)\right]-\frac{\omega\mu _0\sigma_{0}}{2k_{0}D}\int_{-w/2}^{w/2} J_y(x')\, dx'
\end{equation}
Note that, since the external field is now a normally incident plane wave, $E^{ext}_{x}$,$E^{ext}_{y}$ do not depend on $x$.
This simple equation is solved by
\begin{equation}
\label{eq4-ba-sol}
J_y(x)=\sigma_{0}\left[ \frac{E_y^{ext}}{1+\gamma} +\frac{\gamma}{1+\gamma}\frac{I_{x}}{\sigma_{a}w}- \frac{J_x(x)}{\sigma_{a}}\right]
\end{equation}
where
\begin{align}
&I_{x}=\int_{-w/2}^{w/2}J_{x}(x') dx' \label{ix}\\
&\gamma=\frac{\omega\mu _0\sigma_{0}w}{2k_{0}D}=\frac{\eta_{0}\sigma_{0}w}{2D}
\end{align}
where $\eta_{0}=(\mu_{0}/\varepsilon_{0})^{1/2}$.

Next, consider the equation for $J_{x}$.  Upon replacing $G_{0}$ by \eqref{gp} in \eqref{eq3-c}, keeping the leading order term, and using \eqref{eq3-a}, one arrives at
\begin{equation}\label{integ2}
\begin{split}
&\frac{J_{x}(x)}{\sigma_{xx}}= F + \\
&\frac{1}{j\omega \varepsilon _0}\sum_{l=-\infty}^{\infty} \frac{d}{dx} \int_{- w/2}^{w/2} G_0(x - x'-lD) \frac{d J_x(x')}{dx'}dx'
\end{split}
\end{equation}
where
\begin{equation}\label{sx}
F=E_x^{ext}+\frac{\sigma_{xy}}{\sigma_{xx}}\left[ \frac{E_y^{ext}}{1+\gamma} +\frac{\gamma}{1+\gamma}\frac{I_{x}}{\sigma_{a}w}\right]\\ - \frac{\omega \mu _0 I_{x}}{2k_{0}D}
\end{equation}
and we have used 
\begin{equation}
\frac{dG(x)}{dx}=\sum_{l=-\infty}^{\infty} \frac{d G_0(x -lD)}{dx} + O\left( k_{0}D\right)\label{dgdx}
\end{equation}

This equation is identical to the one describing an array of graphene ribbons in the absence of a magnetic field \cite{khavasi2014analytical}, except for $F$ replacing the external field $E_{x}^{ext}$. Using perturbation theory, the solution of \eqref{integ2} can be written as
\begin{align}\label{expan}
&J_{x}(x)=F\sum_{n=1}^{\infty} \mathcal{Y}_{n}S_{n}\psi_{n}(x) \\
 &\mathcal{Y}_{n}=\frac{\sigma_{xx}(2j\omega\varepsilon_{0}/q_{n})}{\sigma_{xx}+2j\omega\varepsilon_{0}/q_{n}}\label{sigman}\\
&S_{n}=
\int\limits_{ - w/2}^{w/2}\psi_{n}(x')dx'\label{sn}
\end{align}
Here $\psi_n(x)$ and $q_{n}$ satisfy the eigenvalue equation\cite{khavasi2014analytical,rahmanzadeh2018adopting}
\begin{equation}
\frac{1}{\pi}\int_{-w/2}^{w/2}\frac{1}{x-x'}\frac{d\psi_{n}(x')}{dx'}dx'=q_{n}\psi_{n}(x)\label{eigv}
\end{equation}
where $\psi_{n}(x)$ is subject to the normalization condition
\begin{equation}
\int_{-w/2}^{w/2} \psi_n^2(x) dx=1
\end{equation}
Note that  $\mathcal{Y}_{n}$ in \eqref{sigman} equals the  admittance  of a capacitance $2\varepsilon_{0}/q_{n}$ in series with a conductance $\sigma_{xx}$.

The constant $F$  self-consistently depends on $J_{x}$ through $I_{x}$ [see \eqref{ix},\eqref{sx}] . To calculate $I_{x}$, an integration is carried out over \eqref{expan}, and \eqref{sx} is used. After basic calculations one obtains
\begin{align}\label{ixf0}
&I_{x}=\frac{Y }{ 1 +\frac{1}{2}\zeta_{0}  Y} DF_{0}\\
&Y=\frac{1}{D}\sum_{m=1}^{\infty}\mathcal{Y}_{m}S_{m}^{2}\label{Ypar}\\
&F_{0} =E_x^{ext}+\frac{\sigma_{xy}}{\sigma_{xx}}\frac{E_y^{ext}}{1+\gamma}\label{f0e}\\
&\zeta_{0}=\eta_0\left[ 1- \frac{\sigma^{2}_{xy}}{\sigma^{2}_{xx}(1+\gamma)}\right]
\end{align}
Substitution in \eqref{sx} then yields
\begin{equation}
F=\frac{F_{0} }{ 1+\frac{1}{2}\zeta_{0}  Y}
\end{equation}

In all the aforementioned equations, the graphene ribbons are assumed to be suspended in free space. In a different medium with a relative dielectric constant of $\varepsilon_{r}$, all the expressions obtains remain valid provided that $\varepsilon_{0}$ and $k_{0}$ are replaced by $\varepsilon_{0}\varepsilon_{r}$ and $k_{0}\sqrt{\varepsilon_{r}}$, respectively. 

\subsection{Scattering parameters}

The distribution of surface current on graphene ribbons is next used to calculate the  reflection and transmission coefficients of the structure. To that end, the electric field in the region $z<0$ is expressed as the Rayleigh expansion
\begin{equation}
\label{eq13} 
\begin{split}   
\left( E_{x}^{ext}\hat{\mathbf x} + E_{y}^{ext}\,\hat{\mathbf y} \right)  &e^{ - j k_{0}z }\\ + 
\sum_{n=-\infty}^{\infty}&\left( E_{x,n}^r\hat{\mathbf x} + E_{y,n}^r\,\hat{\mathbf y} \right) e^{ -j k_{x,n} x + j k_{z,n}z }
\end{split}
\end{equation}
Similarly, the electric field in the region $z>0$ is written as
\begin{equation}
\label{eq13b}    
 \sum_{n=-\infty}^{\infty}\left( E_{x,n}^t\hat{\mathbf x} + E_{y,n}^t\,\hat{\mathbf y} \right) e^{ -j k_{x,n} x - j k_{z,n}z },\, z > 0
\end{equation}
In these relationships  
\begin{equation}
k_{x,n}=2\pi n/D, \, k_{z,n}=\sqrt{k_{0}^{2}-k_{x,n}^{2}}
\end{equation}

In the sub-wavelength regime where  $k_{0}D
\ll 2\pi$ only the zero-order mode ($n=0$) is propagating. All other modes are evanescent. For sub-wavelength ribbon arrays the reflection and transmission coefficient are, therefore, defined using the zero-order mode amplitudes alone, according to the relationships
\begin{align}
\label{eq14} 
&\left[ \begin{array}{c} E_{x,0}^r\\E_{y,0}^r\end{array}\right] = \left[\begin{array}{cc} R_{xx}&R_{xy}\\R_{yx}&R_{yy}\end{array}\right]\,\left[\begin{array}{c}E_x^{ext}\\E_y^{ext}\end{array}\right]\\
&\left[ \begin{array}{c}E_{x,0}^t\\E_{y,0}^t\end{array}\right] = \left[\begin{array}{cc}T_{xx}&T_{xy}\\T_{yx}&T_{yy}\end{array}\right]\,\left[\begin{array}{c}E_x^{ext}\\E_y^{ext}\end{array}\right]
\end{align}
In order to compute the reflection and transmission matrices, the usual boundary conditions on electric and magnetic fields is applied at $z=0$, and an integration is carried out over the resulting equations on a single period, i.e., from $x=0$ to $x=D$. It then follows that
\begin{align}\label{erx}
&E^{r}_{x,0}=E^{t}_{x,0}-E^{ext}_{x}=-\frac{\omega\mu_{0}}{2k_{0}D}I_{x}\\
&E^{r}_{y,0}=E^{t}_{y,0}-E^{ext}_{y}=-\frac{\omega\mu_{0}}{2k_{0}D}I_{y}\label{ery}
\end{align}
where
\begin{equation}
I_{y}=\int_{-w/2}^{w/2}J_{y}(x')dx'
\end{equation}

The next step is to express $I_{x},I_{y}$ in terms of the incident field. To that end, we carry out an integration over \eqref{eq4-ba-sol} from $x=-w/2$ to $x=w/2$. This results in
\begin{equation}\label{eqixy}
I_{y}+\frac{\sigma_{xy}}{\sigma_{xx}}\frac{I_{x}}{1+\gamma}=\frac{\sigma_{0}w}{1+\gamma}E_{y}^{ext}
\end{equation}
where we have used \eqref{eq7}. Combination of \eqref{ixf0},\eqref{f0e}, \eqref{eqixy}, \eqref{erx}- \eqref{eq14} then yields

\begin{align}
\label{Rxx}
&R_{xx}=T_{xx}-1=-\frac{\frac{1}{2}\eta_{0}Y}{1+\frac{1}{2}\zeta_{0}Y}\\
&R_{xy}=-R_{yx}=T_{xy}=-T_{yx}=\frac{\sigma_{xy}R_{xx} }{\sigma_{xx}(1+\gamma)}\label{Rxy}\\
&R_{yy}=T_{yy}-1=-\frac{\gamma}{1+\gamma}-\frac{R_{xy}^{2}}{R_{xx}} \label{Ryy}
\end{align}
For simplification, in all the equations of this subsection, we assume that graphene is suspended in free space. These equations can be readily generalized to the case where the ribbon is sandwiched between two homogenous media with permittivities $\varepsilon_1$ and $\varepsilon_2$, respectively as shown in Fig.\ref{GrapeheneRibbon}. 

\section{RESULT, DISCUSSION AND APPLICATION}
\subsection{Numerical results }
In this section we validate and verify the accuracy of the proposed method through some examples. Consider a periodic array of graphene ribbons in free space with $D=4\mu m, w=2\mu m,E_f=0.5eV,\tau=1ps,B_0=10 T$. For a normally incident plane wave, the magnitude of reflection coefficients  \eqref{Rxx}, \eqref{Rxy} and \eqref{Ryy} is plotted as function of frequency in Fig.2. For the sake of comparison the results obtained using effective medium theory \cite{gomez2016magnetically} (dotted blue), and finite integration technique (FIT) (dashed black) are also plotted. The FIT results are obtained using CST Microwave Studio 2018. Excellent agreement can be observed between the FIT results and our analytical method. By contrast,  the effective medium approach \cite{gomez2016magnetically} can not reproduce the electromagnetic response of magnetically-biased graphene ribbons.

\begin{figure} [t]
\centering
\begin{subfigure}
\centering
\includegraphics[width=\linewidth]{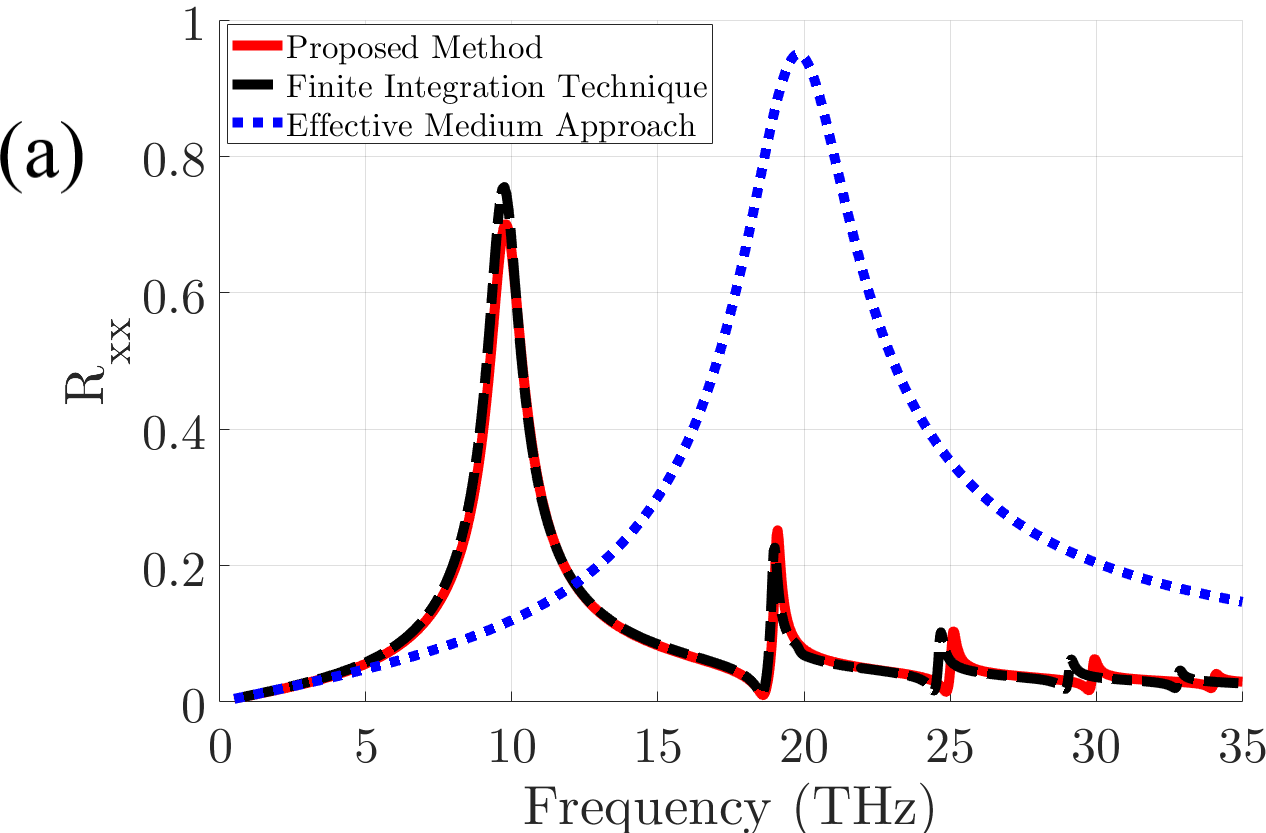}
\label{Fig_RXX}
\end{subfigure}
\begin{subfigure}
\centering
\includegraphics[width=\linewidth]{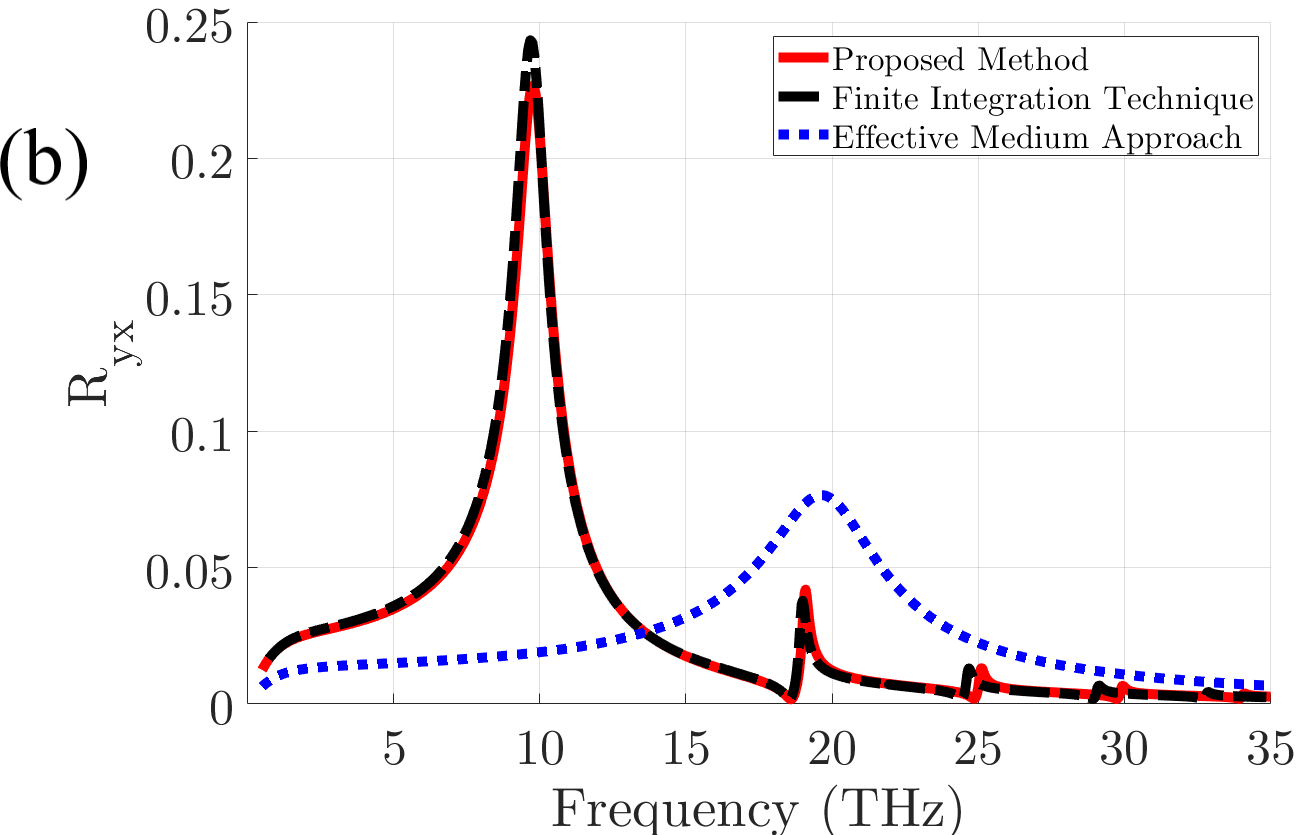}
\label{Fig_Ryx}
\end{subfigure}
\begin{subfigure}
\centering
\includegraphics[width=\linewidth]{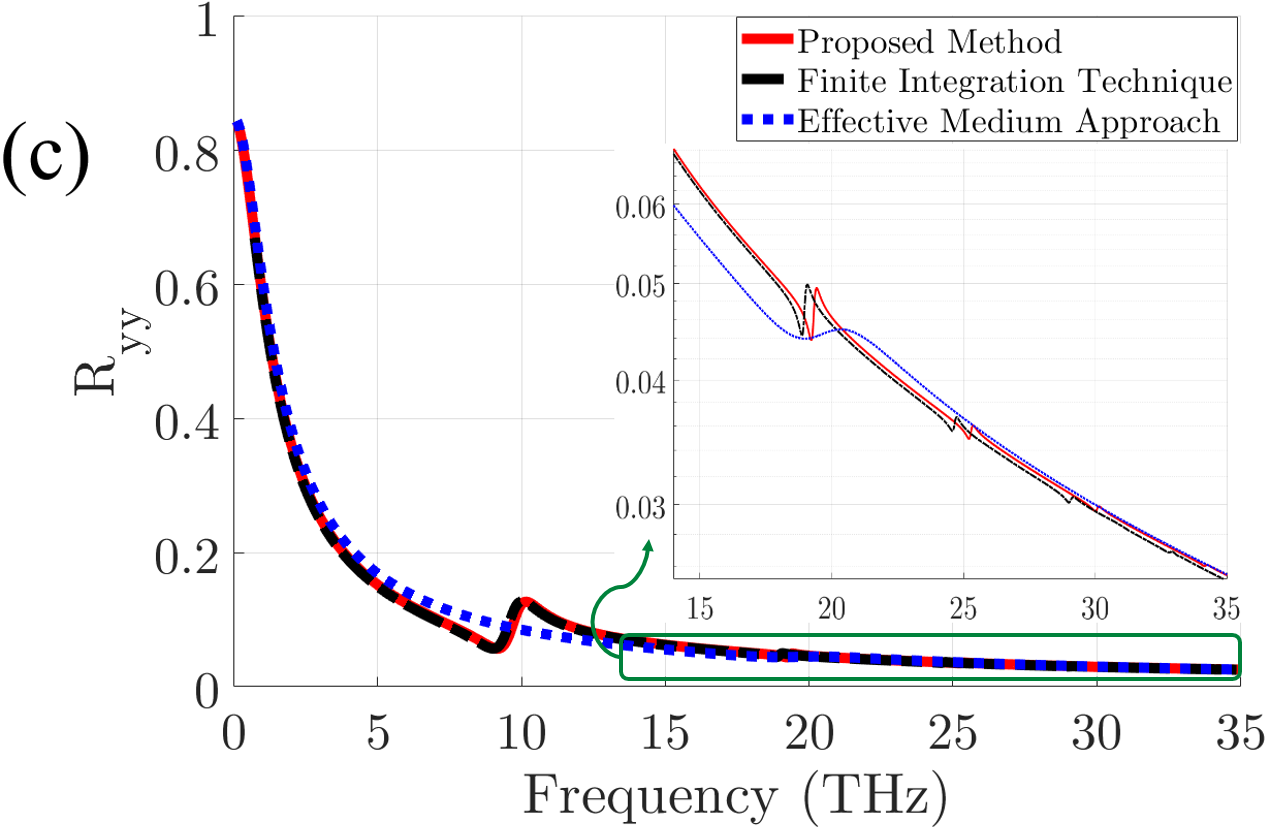}
\label{Fig_Ryy}
\end{subfigure}
\caption{(a)$R_{xx}$ and (b) $R_{yx}$  (c) $R_{yy}$of a periodic array of magnetically-biased graphene ribbons. The graphene ribbons parameters are assumed as  $D=4\mu m, w=2\mu m,E_f=0.5eV,\tau=1ps,B_0=10 T$}
\label{NumericalExample}
\end{figure}
The resonance features observed in Fig.\ref{NumericalExample} can be explained as follows. Equations \eqref{expan},\eqref{sigman} imply that when $q_n\sigma_{xx}+2j\omega\varepsilon_0$ becomes very small  for some $n$, $\mathcal{Y}_{n}$ becomes large and a resonance occurs. Near the corresponding resonance frequency, the distribution of $J_{x}(x)$ is predominantly given by $\psi_{n}(x)$.  Apart from constant terms, the distribution of $J_{y}(x)$ is identical to $J_{x}(x)$ as can be seen from \eqref{eq4-ba-sol}. Fig.\ 3 shows the magnitude of $J_x,J
_y$ across a graphene ribbon at the first ($n=1$) and second ($n=3$) resonances, that occur at 9.78- and 19.13 THz, respectively. It must be noted that $\psi_{n}(x)$ is an odd function of $x$ with respect to $x=0$ (center of a ribbon) for even values of $n$. As a result $S_{n}=0$ for even values of $n$  [see \eqref{sn}] so that only odd values of $n$ enter the summation \eqref{expan}. The resonances observed are thus due to odd modes $n=1,3,\cdots$.  Near these resonance frequencies, peaks in reflection coefficients are observed. Note, however, that $R_{yy}$ does not show any resonance features in the absence of an external magnetic bias, i.e., when $\sigma_{xy}=0$. This can be seen from (\eqref{Ryy}). 

\begin{figure}
\centering
\includegraphics[width=\linewidth]{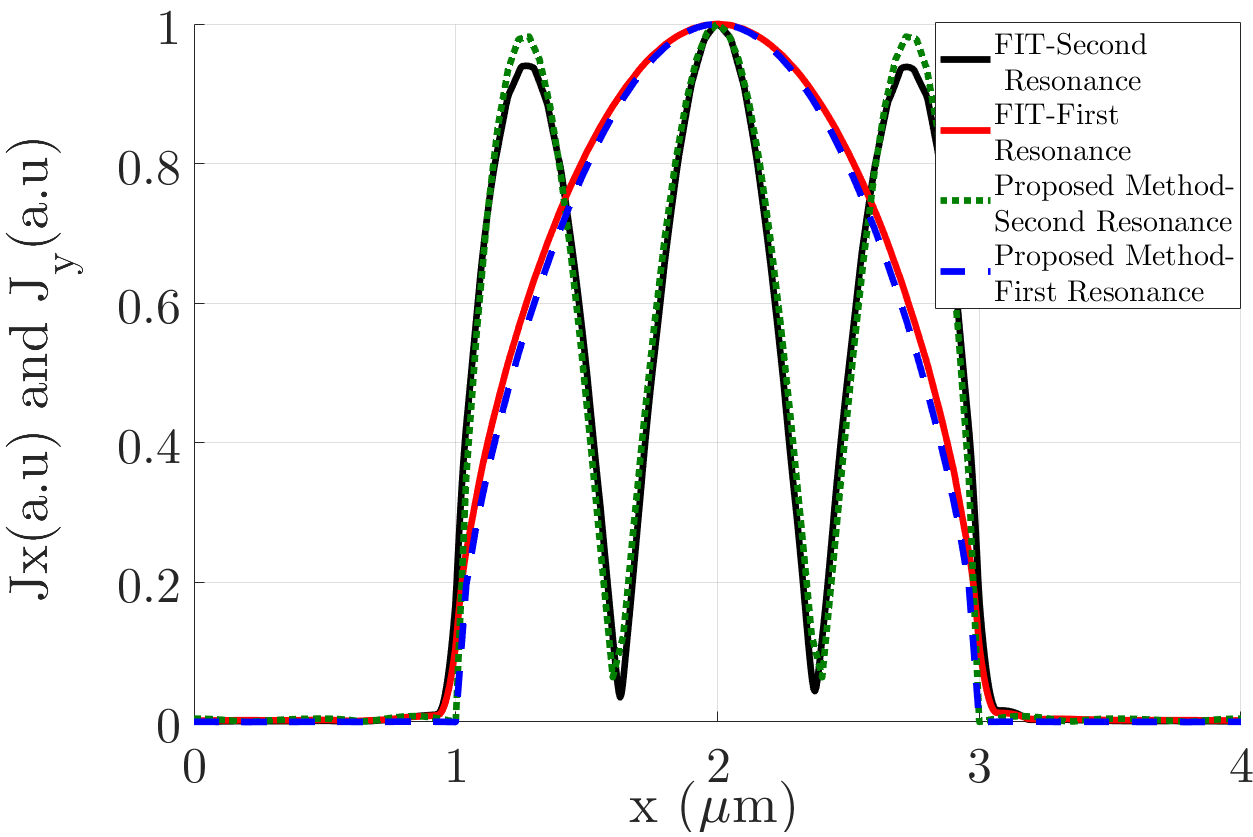}
\caption{Current distribution on the graphene ribbon at the vicinity of the first two resonances}\label{Surface Current}
\end{figure}

\begin{figure}
\centering
\includegraphics[width=\linewidth]{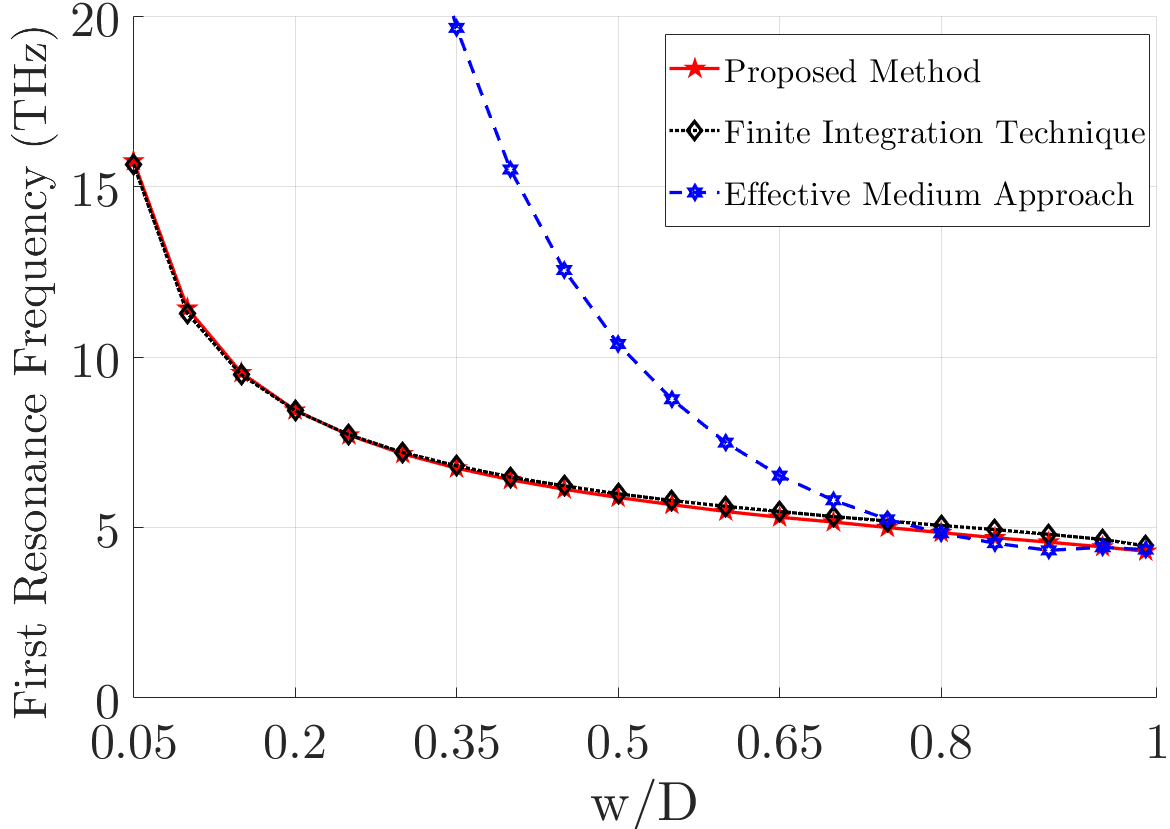}
\caption{First resonance frequency of $R_{xx}$ for different filling factors $w/D$ ($B_0=7.5T,\tau=1ps$ and  $E_f=0.3ev$) }
\label{SweepFig}
\end{figure}

With regards to the limitations of the presented theory, it should be noted that our analysis assumed operation in the sub-wavelength regime where $k_{0}D\ll 1$.  As shown in Fig.\ref{NumericalExample}, when the frequency increases, the accuracy of the proposed method decreases. For the position of the first resonance, we have a relative error of $0.5\%$ while the relative error for the fourth resonance is $2.71\%$. Moreover, solution \eqref{expan} is based on perturbation theory in which the interaction between neighboring ribbons is assumed not to affect the eigenvalue equation \eqref{eigv} \cite{khavasi2014analytical}. Increasing the fill factor ($w/D$), will lead to a smaller gap and  a stronger interaction between neighboring ribbons. One would, therefore, expect the accuracy of our method to decrease.  The frequency of the first resonance is plotted as function of fill factor in Fig.\ref{SweepFig} for  $D=10\mu m, E_f=0.3eV,\tau=1ps,B_0=7.5 T$. It is observed that by increase in the fill factor, the relative error in predicting the resonance frequency is slightly increased.

Finally, derivation of \eqref{eq4-b-sol} was based on \eqref{asy}. According to \eqref{eq2} and \eqref{eq7},  when $E_f$ is large, \eqref{asy} cannot be satisfied. If the Fermi level energy increases, the accuracy of the proposed method can slightly decrease. The effect of this approximation, as well as perturbation theory, is small compared to the   sub-wavelength approximation. The latter, however, does not pose a strict limitation, since anisotropic graphene ribbons have almost always been utilized in the sub-wavelength regime \cite{poumirol2017electrically,fallahi2012manipulation,tymchenko2013faraday,gomez2016magnetically}. Consequently, the proposed analysis is much more accurate and affordable than its counterparts already reported in the literature \cite{tymchenko2013faraday,feizi2018modeling,gomez2016magnetically}.

\subsection{Design of a Faraday rotator }
\begin{figure}
\centering
\begin{subfigure}
\centering
\includegraphics[width=\linewidth]{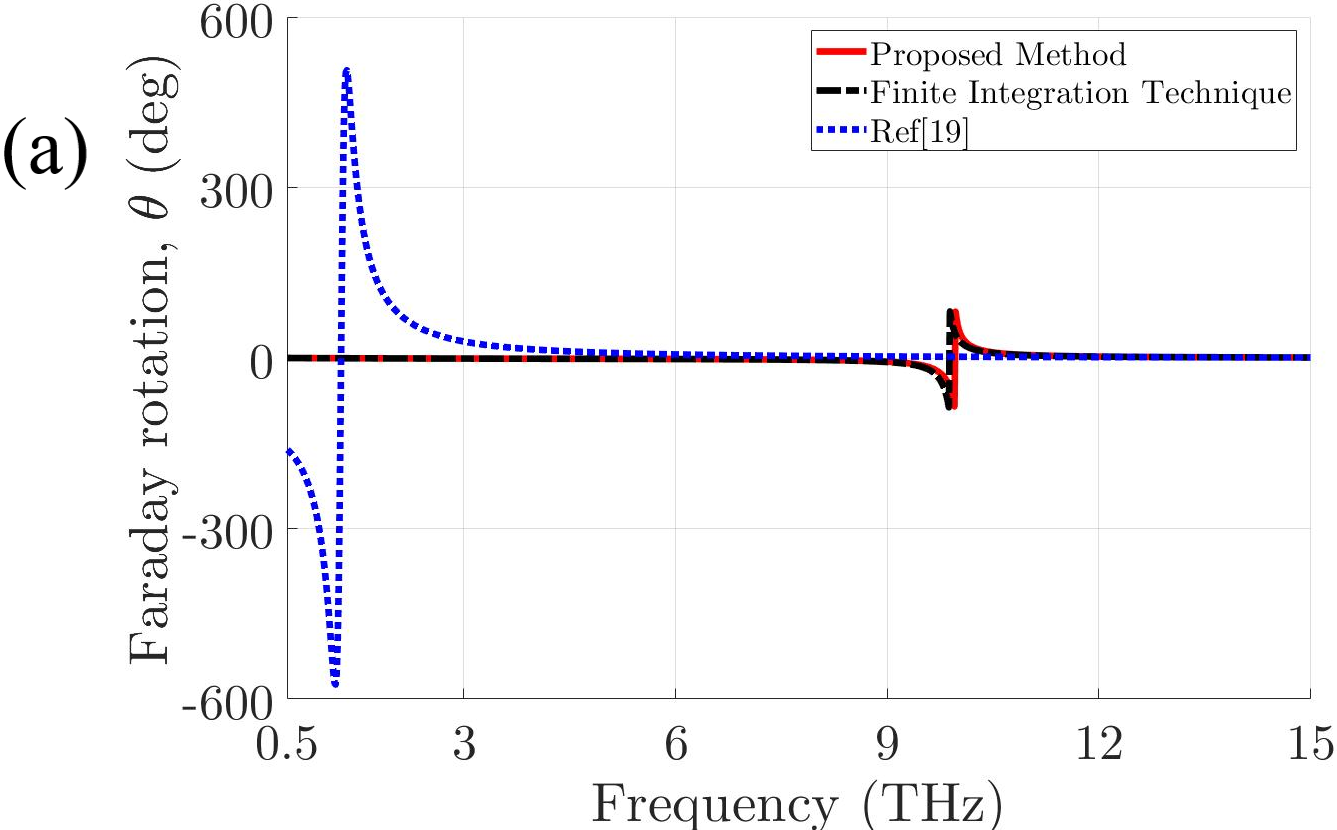}
\label{FR_Example}
\end{subfigure}
\begin{subfigure}
\centering
\includegraphics[width=\linewidth]{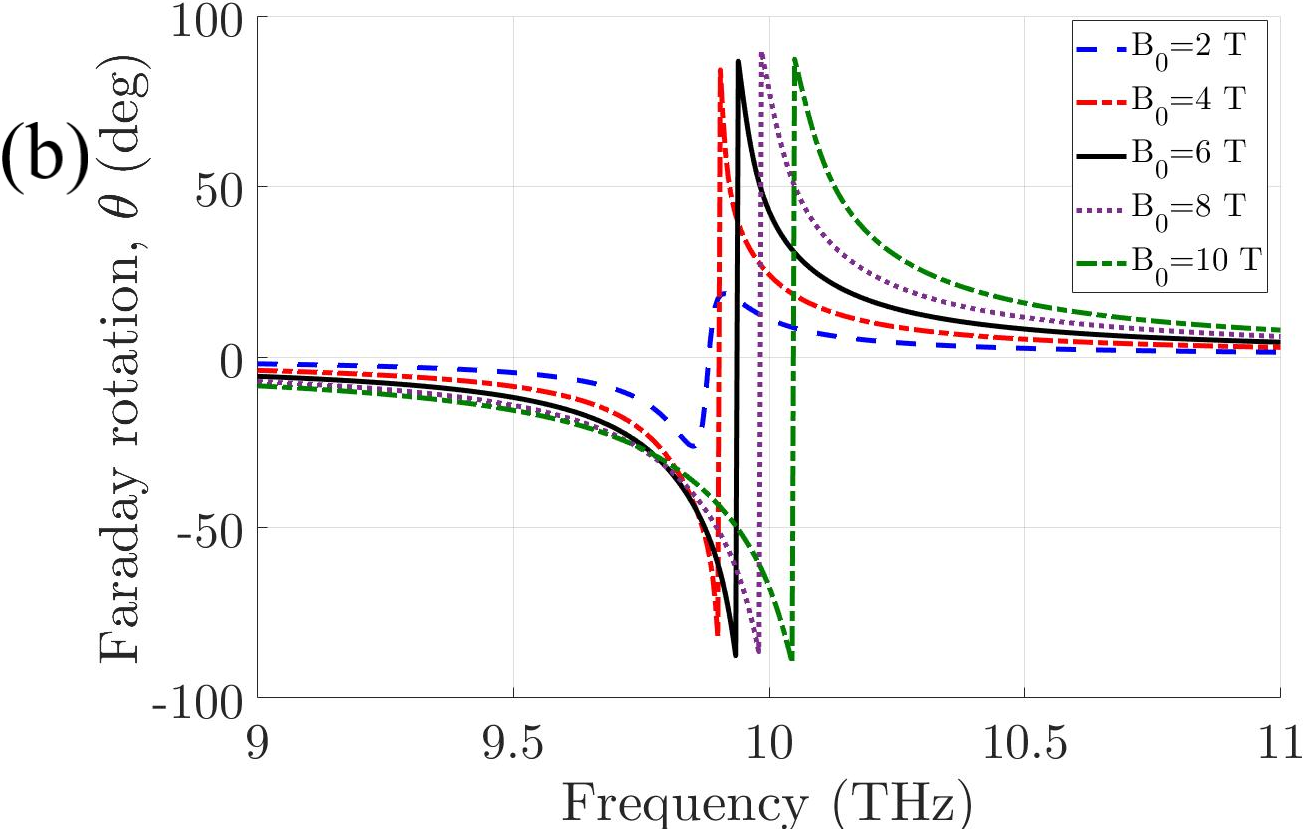}
\label{SweepB0}
\end{subfigure}
\caption{Faraday rotation angle at various magnetic field bias for graphene ribbons.structure parameter are designed as  $D=4.5\mu m, w=2.7\mu m,E_f=0.8eV,\tau=2ps$}
\label{FaradyRotatorFig}
\end{figure}

Faraday rotation is one of the most attractive phenomena in optics, with many applications in optical diodes, sensing, magnetic microscopy, optical communications, data storage, optical isolators, and phase modulators \cite{crassee2011giant,tymchenko2013faraday}. For the array of graphene ribbons under magnetic bias, Faraday rotation angle can be directly computed as follows\cite{tymchenko2013faraday}:
\begin{equation}
 \label{Farady_Rotation} 
{\theta _F} = \frac{1}{2}\mathrm{arg}\left( \frac{{{T_{xx}} - j\,{T_{yx}}}}{{{T_{xx}} + j\,{T_{yx}}}}\right)
\end{equation}
where $T_{xx}$ and $T_{yx}$ can be calculated from \eqref{Rxx} and \eqref{Rxy}. Large Faraday rotation angles are observed at resonance.  To achieve a giant Faraday rotation at 10 THz, the structure parameters are designed as  $D=4.5\mu m, w=2.7\mu m,E_f=0.8eV,\tau=2ps,B_0=7T$. The  results are shown in Fig.\ref{FaradyRotatorFig}. A good agreement between the results of our method and numerical FIT is observed. Also, we compare our results with the semi-analytical results \cite{tymchenko2013faraday}. It is clear that our method yields much more accurate results for the Faraday rotation angle of the magnetically-biased graphene ribbons. This is because the analysis in \cite{tymchenko2013faraday} is based on the low-relaxation-time assumption. Besides, it assumes that period $D$ is much smaller than the free-space wavelength.  Neither condition is satisfied here. 

Finally, we investigate the effect of magnetic bias on performance of the Faraday rotator in Fig.\ref{FaradyRotatorFig}b. It can be seen that increasing the  magnetic bias leads to larger  Faraday rotation angles.  Also,  giant Faraday rotation occurs at  higher frequencies.

\section{Conclusion}
 Based on integral equations governing the field and the surface current density on magnetically-biased graphene ribbons, this paper proposed a analytical, fast, and accurate method for the analysis of such structures. Unlike previous studies, the proposed method does not suffer from extremely slow convergence rates of numerical methods, and more importantly is able to predict spectral resonance features, not possible with previous analytic solutions. The expressions were derived under a quasi-static approximation and extended to periodic arrays of graphene ribbons using perturbation theory. By Rayleigh expansion, we determined the reflection/transmission coefficient using a rigorous method. The proposed method was validated against commercial software equipped with finite integration technique (FIT), showing excellent agreement between the two solutions. In addition, a Faraday rotator was designed and investigated as an important application. The present study suggests a new method for the analysis of magnetically-biased graphene ribbons, which is more precise, reliable, and affordable than the methods already reported in the literature. The proposed method might also be useful for the analysis and development of different devices such as polarizer, isolators, and so on.

\appendices
\setcounter{equation}{0}
\numberwithin{equation}{section}

\setcounter{figure}{0}
\numberwithin{figure}{section}

\section{\hl{}}
\label{Appendix1}
Consider the Green function \eqref{phank}.  The Hankel function is written as
\begin{equation}
\label{A1}
 H_0^{(2)}\left( z \right) =  - \frac{1}{\pi }\int\limits_{\pi  + j\infty }^{ - j\infty } e^{ - jz\sin \theta } d\theta 
\end{equation}
were the integration is carried out over the path shown  on page 916 of \cite{gradshteyn2014table} in the complex $\theta$ plane. 
Since $\left| x \right| < D $, one has: 
\begin{equation}
\label{A3}
\begin{split}
G(x)&=-\frac{1}{4\pi j}\int\limits_{\pi  + j\infty }^{ - j\infty }
e^{jk_{0}x\sin\theta}\sum_{l=-\infty}^{\infty} e^{ - jk_{0}lD\sin \theta }d\theta \\ &=\frac{1}{4\pi }
\int\limits_{\pi  + j\infty }^{ - j\infty }
\frac{\cos\left[ \left( \frac{1}{2}D-|x|\right) k_{0}\sin\theta\right] }{\sin\left(\frac{1}{2}k_{0}D\sin\theta \right) } d\theta
\end{split}
\end{equation}
where it is assumed that the integration contour does not pass through points where  $\sin(\theta)=0$.  

\eqref{A3} can be rewritten as
\begin{equation}\label{A4}
\begin{split}
&G(x)=\frac{1}{2jk_{0}D}+\\ &\frac{1}{4\pi }
\int\limits_{\pi  + j\infty }^{ - j\infty }\left(
\frac{\cos\left[ \left( \frac{1}{2}D-|x|\right) k_{0}\sin\theta\right] }{\sin\left(\frac{1}{2}k_{0}D\sin\theta \right) } -\frac{2}{k_{0}D\sin\theta}\right)d\theta
\end{split}
\end{equation}
where we have used
 \begin{equation}
\label{A6} 
 \int\limits_{\pi  + j\infty }^{ - j\infty } {\frac{{d\theta }}{{\sin \theta }}}  =   -j\pi
\end{equation}
The integrand in \eqref{A4} is devoid of singularities arising at $\theta =0,\pi$ so that the integration contour is deformed into the one shown on page 916 of \cite{gradshteyn2014table}. The resulting integral is written as
\begin{equation}
\label{A4a}
\begin{split}
&-\frac{1}{4\pi }
\int\limits_{0 }^{ \pi }\left(
\frac{\cos\left[ \left( \frac{1}{2}D-|x|\right) k_{0}\sin\theta\right] }{\sin\left(\frac{1}{2}k_{0}D\sin\theta \right) } -\frac{2}{k_{0}D\sin\theta}\right)d\theta \\
&+\frac{1}{2\pi }
\int\limits_{0 }^{ \infty }\left(
\frac{\cosh\left[ \left( \frac{1}{2}D-|x|\right) k_{0}\sinh t \right] }{\sinh \left(\frac{1}{2}k_{0}D\sinh t \right) } -\frac{2}{k_{0}D\sinh t}\right)dt
\end{split}
\end{equation}
Since $k_0D$ is small (and so is $k_0 x$) the integrand in the first integral can be expanded in powers of $k_{0}D$,$k_{0}|x|$. The result can be easily shown to be of the order $k_{0}D$.  To compute the second integral it is rewritten as
\begin{equation}
\label{A4b}
\begin{split}
\frac{1}{2\pi }
\int\limits_{0 }^{ \infty }\left[
\frac{\cosh\left( \alpha u \right) }{\sinh u } -\frac{1}{u}\right]\frac{du}{\sqrt{u^{2}+\frac{1}{4}\left( k_{0}D\right)^{2}}}=\\
\frac{1}{2\pi }
\int\limits_{0 }^{ \infty }\left[
\frac{\cosh\left( \alpha u \right) }{\sinh u } -\frac{1}{u}\right]\frac{du}{u}+O\left( k_{0}D\right)^{2}
\end{split}
\end{equation}
where $\alpha=1-2|x|/D$ and $u=(k_{0}D/2)\sinh t$. Inserting this result in \eqref{A4} leads to \eqref{gp}. Take note that
\begin{equation}
\label{phank3}
\begin{split}
&\frac{dG(x)}{dx} =-\frac{ x}{\pi D |x|  }
\int\limits_{0 }^{ \infty }
\frac{\sinh\left( \alpha u \right) }{\sinh u }du +O\left( k_{0}D\right)\\
&= -\frac{ \pi x}{ D|x| } \cot \left( \frac{\pi |x| }{D}\right)+O\left( k_{0}D\right)
\\ &
= -\frac{1}{2\pi} \sum_{n=-\infty}^{\infty}\frac{1}{x-nD}+ O\left( k_{0}D\right)
\end{split}
\end{equation}
which is identical to \eqref{dgdx}.
\bibliographystyle{IEEEtran}
\bibliography{IEEEabrv,Bibliography}
\vspace{0 cm}
\begin{IEEEbiographynophoto}{Mahdi Rahmanzadeh}
 was born in Amol, Iran, in 1994. He received the B.Sc. degree from the Babol Noshirvani Institute of Technology, Babol, Iran and the M.Sc. degree from the Iran University Science and Technology, Tehran, Iran all in electrical engineering in 2015 and 2018 respectively . He is currently pursuing the Ph.D. degree in electrical engineering with the Department of Electrical Engineering, Sharif University of Technology. His current research interests include Electromagnetics,metamaterials, graphene metasurfaces and photonics.
\end{IEEEbiographynophoto}
\begin{IEEEbiographynophoto}{Behzad Rejaei}
 received the M.Sc. degree in electrical engineering from the Delft University of Technology, Delft, The Netherlands, in 1990, and the Ph.D. degree in theoretical condensed matter physics from the University of Leiden, Leiden, The Netherlands, in 1994. From 1995 to 1997, he was a member with the Physics Faculty, Delft University of Technology, where he carried out research on mesoscopic charge-density-wave systems. Between 1997 and 2010, he was with the Department of Electrical Engineering, Mathematics, and Computer Science, Delft University of Technology. He is currently an Associate Professor with the Department of Electrical Engineering, Sharif University of Technology, Tehran, Iran. His research interests are in the areas of electromagnetic modeling of integrated passive components, passive magnetic devices, and graphene plasmonics.
\end{IEEEbiographynophoto}
\begin{IEEEbiographynophoto}{Mohammad Memarian}
(S’08–M’16) received the B.A.Sc. (Hons.) and M.A.Sc. degrees in electrical engineering from the University of Waterloo, Waterloo, ON, Canada, in 2007 and 2009, respectively, and the Ph.D. degree in electrical engineering from the University of Toronto, Toronto, ON, in 2015.,He was a Post-Doctoral Fellow with the University of California at Los Angeles, Los Angeles, CA, USA. He is currently an Assistant professor with the Sharif University of Technology, Tehran, Iran. His current research interests include electromagnetics, microwaves and antennas, metamaterials, photonic crystals, gratings, metasurfaces, and dielectric resonator filters and system.
\end{IEEEbiographynophoto}
\begin{IEEEbiographynophoto}{Amin Khavasi}
was born in Zanjan, Iran, in 1984. He received the B.Sc., M.Sc., and Ph.D. degrees in electrical engineering from the Sharif University of Technology, Tehran, Iran, in 2006, 2008, and 2012, respectively. Since 2013, he has been a Faculty Member with the Department of Electrical Engineering, Sharif University of Technology, where he is currently an Associate Professor. His current research interests include photonics, the circuit modeling of photonic structures, and computational electromagnetics.
\end{IEEEbiographynophoto}
\vfill
\end{document}